*Biography*

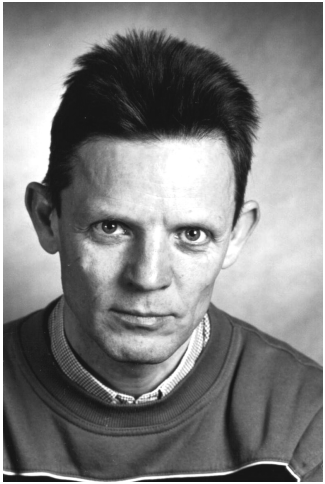

**Dr. Attila Grandpierre** is currently a senior scientific researcher at the Konkoly Observatory of Hungarian Academy of Sciences. He obtained his degree of University Doctor in 1977 from the Eötvös Lóránd University, his academic degree candidate of physical sciences in 1984 from the Hungarian Academy of Sciences and continued his research at the Konkoly Observatory.

Dr. Grandpierre's scientific interests are in the areas of: origin of solar activity, the Sun as a self-organizing system and as a system showing lifelike activities, non-protein based life forms in the Universe, first principles of physics and biology.

# 28

# Fundamental Complexity Measures of Life

**ATTILA GRANDPIERRE**

*Konkoly Observatory of the Hungarian Academy of Sciences*
*H-1525 Budapest, P. O. Box 67, Hungary*
*grandp@iif.hu*

## 1. Introduction

At present, there is a great deal of confusion regarding complexity and its measures (reviews on complexity measures are found in, e.g. Lloyd, 2001 and Shalizi, 2006 and more references therein). Moreover, there is also confusion regarding the nature of life. In this situation, it seems the task of determining the fundamental complexity measures of life is especially difficult. Yet this task is just part of a greater task: obtaining substantial insights into the nature of biological evolution. We think that without a firm quantitative basis characterizing the most fundamental aspects of life, it is impossible to overcome the confusion so as to clarify the nature of biological evolution. The approach we present here offers such quantitative measures of complexity characterizing biological organization and, as we will see, evolution.

Fortunately, some important complexity measures are already established. Two such fundamental complexity measures are the algorithmic complexity of the human brain and the genetic complexity of the human organism. Let us consider how they are obtained.

The complexity measure $C$ of a system consisting from $N$ elements can be characterized by the number of distinctive connections (not including replicas) between the elements $C = Nc_d$ where $c_d$ is the average number of distinctive connections per element (Denbigh, 1975, 99). The term 'distinctive' is necessary, because an airplane is more complex than a watch, but a hundred watches of the same kind are not more complex than a single watch. The complexity measure is stated in terms of information units that estimate the information content $i$ of one element (having an average number of connections) as $I = Ci$. On this basis, the brain's complexity is usually considered in terms of neurons and synaptic connections, e.g. Stripling (2004). For a number of neurons $N_{neurons} = 10^{11}$–$10^{13}$ (Smith, 1997, 921), taking a value for the number of their interconnections as a few thousand per neuron $c_d \sim 10^4$ (Koch and Laurent, 1999), we obtain for the measure of the brain's complexity the number $C_1 = cN_{neurons} = 10^{15}$–$10^{17}$. We emphasize that the complexity measure $C$ measures the number of (distinctive) interconnections, and so it is a dimensionless number that corresponds to objective reality. The general view assumes that a connection (synapse) represents $i \sim 1$ bits of information. In this way, we obtain for the information measure of the brain's complexity a value of $I_1 = I$(human brain) $\sim 10^{15}$–$10^{17}$ bits. Now since algorithmic complexity may be characterized by the size of the memory, we obtain that $I_1 = I_{algorithmic}$(brain).

Although it may seem that this estimation is overly simplified and evident, we point out a baffling problem that surfaces immediately when comparing $I_{algorithmic}$(brain) with the genetic complexity of the whole human organism. Maynard Smith (2000) noted that the genetic information content of human DNA corresponds to "instructions," and assigns a value of $I_2$(DNA) $\sim 10^9$ bits to this information measure. (Notice that pure genetic complexity can be measured by a dimensionless number, characterized by the number of coding base pairs. See Maynard Smith and Szathmáry, 1995, 5). Now it is generally accepted that the DNA controls all the biochemical processes of the organism (e.g. Woski and Smith, 2002, 28). We point out that the comparison of these two firm complexity measures presents a fundamental paradox: How can it be that the genetic complexity of the human organism (including the brain) — $I_2 \sim 10^9$ bits — is smaller than the algorithmic complexity of the human brain, $I_1 \sim 10^{15}$–$10^{17}$ bits? We have to think that the organism as a whole has to be more complex than one of its parts, the brain[1].

---

[1]Actually, a part of brain's activity is governed by self-conscious activity. But its is easy to show that $\dot{I}$(self-conscious activity) $< 100$ bits s$^{-1}$ (we cannot read more than a few pages per minute; Breuer, 1995), and so the information obtained by self-conscious activity during a lifetime $I$ (self-conscious activity per lifetime) $< 10^{11}$ bits dwarfs in

We propose that the solution of this *brain–DNA paradox* lies in the fact that genetic complexity is related to elementary instructions determining the organization of simple biochemical reactions of the cells, while algorithmic complexity corresponds to simple biochemical reactions organized together into cycles or units of reaction sequences. Genetic complexity controls all the biochemical reactions occurring in cells; and so the gross activity of the brain corresponds to a comparatively small subset of biochemical reactions only, which have a much lower level of complexity than that of the algorithmic complexity of the human brain. In this simple model, the brain works only on some special characteristics of neural cells, such as their firing. Therefore, the resolution of the brain–DNA paradox lies in the recognition that genetic complexity corresponds to a deeper level of complexity than the algorithmic complexity of the brain, corresponding to the memory of the neural network.

We note that we are faced immediately with another fundamental problem. The DNA contains $10^9$ bits of information in the sequence of the base pairs, and this is a static form of information. In comparison, the biochemical reactions of cells represent a dynamic flow of information corresponding to their selection, coordination, and timing. If DNA is static, it could not govern the continuously changing reactions. No machine can work without moving components. This is the *DNA–dynamism paradox*. It turns out that with the introduction of the two most firmly established biological complexity measures, we are faced with two paradoxes immediately. In this chapter, we will suggest a simple resolution of these paradoxes.

## 2. Fundamental Complexity Measures of Life

In mathematics, there are no numbers without mathematical rules determining their interactions. Actually, numbers and mathematical rules are fundamentally different entities. In the phenomenological aspect, the most fundamental entities are numbers, sets, and complex sets of sets. In geometry, the most fundamental entities are the point, the geometrical structure (circle, square, etc.), and complex structures comprised of structures of structures. In the organizational aspect, the most fundamental entities are relations, rules, and axioms.

---

comparison to the information represented by the whole neural network $I_I \sim 10^{15}$–$10^{17}$ bits.

We note that the point and the circle have not only different levels of complexity, but they also express different kinds of complexities. Any circle can be represented as consisting from an infinite number of points, and as the correspondence between the points. This idea can be formulated by a simple concept: A circle consists of a set of points in a plane that are equidistant from a single point which is the centre of the circle. If one were to assign one bit (one yes or no question and its answer) to one point, the information content of the circle would be infinite. In a world in which only points exist, the construction of a circle would require giving the position of the "next" point, consecutively; and so the circle would require an infinite number of yes or no questions and answers, and so an infinite number of bits.

Actually, in the process of constructing the circle from points one by one, still one would need concepts representing deeper level complexities, such as "continuity" and "closed line," which cannot be expressed directly in the language of points. Yet we recognize that the idea of the circle can be characterized by a deeper-level complexity — *algorithmic complexity*. The algorithmic complexity of a circle is a finite and small quantity. The distinction between the two levels of complexities of the circle — it's phenomenal and algorithmic complexity — is fundamental.

In physics, there are no elementary particles without laws of interaction. Elementary particles form structures, and complex structures of structures. Their interactions are described by physical laws. And all the fundamental laws of physical interactions can be derived from the least action principle.

The next step towards finding quantitative measures of biological organization is to consider what a machine is. A machine is a special arrangement of service parts or components put together according to its blueprint. In the organizational aspect, the machine is made up by the assembly of its components, and by the realization of its blueprint, which determines the interactions between its components. The components have to be produced, and the instructions required to their production represent a type of algorithmic complexity: The blueprint determines the relation of the components to each other. Therefore, the specified complexity of the blueprint corresponds to a still deeper level of complexity than the complexity of the production of its components. A machine is governed by its blueprint. Yet not only the blueprint, but the components themselves are static, passive. In the phenomenological aspect, regarding their complexity levels, machines consist of: a) particles, b) components, and c) components put together.

In comparison, living organisms consist first of all of processes of biological interactions. Their complexity levels are a) simple biochemical reactions, b) biochemical cycles and units of reaction sequences, and, at their deepest organizational level, c) biological organization. Ultimately, living organisms are governed by genetic instructions determining which reactions and units of reaction sequences have to occur, and when and where. Genetic instructions change from one timestep to the next, and involve all the components of a living organism simultaneously. This is a fundamental difference between living organisms and machines.

Living organisms are creative beings, as evolution, plants sciences and ethnology show. Regarding their physical level, living organisms can be compared to robots. Living organisms can behave in the next timestep as a new robot, a robot with a new function. They are able to invent new blueprints serving new, unforeseen tasks. Living organisms are dynamic at their deepest level of complexity.

The difference between the machine and the living organism is like the difference between numbers and mathematical rules. The complexity of the machine is phenomenological, static, and passive; while that of the living organism is organizational, dynamic, and active.

Let us approach the distinctions between machines and living organisms in light of the difference between physical "organization" (termed as "self-organization") and biological organization. As the root of the word "organization" ("organ") tells, organization belongs to the realm of biology. Physical "organization" is present in the order of crystals, of magnets, of snowflakes patterns, of convection patterns, of reaction-diffusion patterns, etc. Physical "organization" represents actually not organizational, but ordering processes.

Actually, ordering and organization are two fundamentally different processes (see e.g. Denbigh, 1975, 89-98; Elitzur, 1994). When starting from the living state, the larger is the order, the smaller is the organization, as shown when a living organism becomes frozen. In physical ordering, patterns of elements can be generated, and in man-made machines they follow prescribed rules. In living organisms, biological organization generates new rules from time-step to time-step. *Biological processes are governed by complexity* present in the boundary conditions of physical equations. In contrast, even the most complex *physical orderings* (like reaction-diffusion processes) *are governed by physical equations*, and their boundary conditions are simple. Actually, biological processes are governed by an extraordinarily complex system of time-dependent boundary conditions of the physical equations.

Written sentences are composed of words, and words are composed of letters. Letters represent the fundamental elements or building blocks, corresponding to numbers in mathematics, and particles in physics. Words correspond to structures in mathematics, and patterns in physics. Sentences correspond to machines and organisms. One can compare the difference between machines and organisms to the difference between syntax and semantics. Machines follow the (once-for-all) established syntactical rules only, since physics does not deal with problems of meaning. Sentences written by a computer following merely the rules of syntax will form an incoherent sequence of sentences, most of which will be without any meaning. In contrast, when the sentences follow meaning, the result will be a poem, a novel, or a book on science: one single organism.

A living organism follows the continuously changing internal and external contexts, and reacts to them on the basis of its own principle driving its biological organization towards the optimization of life's conditions. Biological organization is like writing, while physical ordering is mechanical repetition of words following merely syntactical rules, if any. It is these syntactical rules that represent algorithmic complexity. In contrast, the semiotic principles correspond to a deeper, principal level of complexity. This is why machines cannot rebuild themselves from time-step to time-step. At the same time, this is the most fundamental property of organisms.

Denbigh (1975, 96-97) emphasizes that "one cannot speak of an entity as being organized without at once raising the question: What is it organized for? (…) A machine is not explainable by the laws of physics and chemistry (even though the material of which it is composed obeys these laws); machines have always to be understood in terms of their own specific operational principles laid down by those who design them."

The chemical reactions within living organisms are mostly organized into cycles and units of reaction sequences and thus represent instances of algorithmic complexity. Genetic instructions elicit the proper cycle and unit of reaction sequences at the proper place and time. They correspond to a deeper level of complexity than the algorithmic complexity of biochemical cycles and units of reaction sequences. The ultimate level of biological organization occurs at the genetic level. Genetic complexity characterizes a deeper level of complexity than algorithmic complexity. In every timestep, biological organization switches into new algorithms and new contexts. The new perspectives are determined by the new internal and external factors.

Ontologically, there are three fundamental levels of complexity corresponding to the fundamental levels of existence. We observe by our outer senses the phenomenal world $W_0$. $W_0$ is governed by the laws of Nature, corresponding to a deeper ontological level we shall call the 'level of laws' or 'lawlike level' — $W_1$. The laws of Nature do not represent the ultimate level of existence, for they are derived from first principles like the action principle of physics and the Bauer principle (Bauer, 1935/1967, 51) of biology. *By the term "first principle" we mean a principle from which all the fundamental laws of natural sciences (physics, biology, and psychology) can be derived.* Therefore, it appears only three first principles are necessary in the natural sciences: one in physics (this is the action principle), one in biology (this is the Bauer principle; Bauer, 1935/1967), and one in psychology (formulated by Grandpierre, 2005, 76-102).

The Bauer principle states that "*The living and only the living systems are never in equilibrium and they continuously invest work from their free energy resources against the equilibrium that should be reached on the basis of the given initial state and the physico-chemical laws.*" Thus the first principle of biology acts to generate consecutively the initial and boundary conditions that will be the input elements of the physical equations in the next timestep, which in turn correspond to evolving elementary biochemical reactions within the cells. Therefore the phenomenal–algorithmic–genetic complexities correspond to the phenomenal–lawlike–principal ontological levels. These three ontological levels correspond to three complexity levels. We note that the three levels of complexity as interpreted here show a certain similarity to Maynard Smith's (2000) three complexity levels: the morphological, the selection, and the genetic level.

Denbigh (1975, 93) notes that biological organization is basically "the organization of chemical processes each taking place continuously". Now let us estimate the complexity of biological organization of the human organism as expressed in elementary biochemical processes realized within cells. Certainly, the number of chemical reactions per second is larger than the number of ATP molecules produced per second. Kornberg (1989, 65) determined that the average daily intake of about 2500 kcal, corresponding to approximately 100 W, translates into a turnover of a whopping 180 kg of ATP. This number translates into $N_2 = N_{ATP}$(organism) $\sim 2\times10^{21}$ ATP molecule production per second in the human body, or $4\times10^7$ ATP molecule production per cell per second. Regarding the fact that ATP is produced in a chain of electron transfer events, and acts through energy coupling that involves the coupling of two reactions occurring at the same time, at the same place, typically utilizing the same enzyme complex, we find it plausible to assume that the rate of ATP production of $N_{ATP}$(organism) $\sim 2\times10^{21}$ reactions per second is smaller than the number of all chemical reactions of the human organism, $N_3 = N_{reactions}$(organism) $> 2\times10^{21}$ chemical reactions per second. It is clear that both the production of each ATP molecule together with its reactants has to be timed so that the energy coupling can be effective, and that this timing is not completely pre-programmed because it depends on dynamic, on-going changes of state at the cellular, intercellular, and global organizational levels. One may presume that at least 1 bit is necessary for the proper timing of a chemical reaction. Therefore the flux of biochemical reactions would correspond to a rate of information production $\dot{I}_1 = \dot{I}_{biochem} > 2\times10^{21}$ bits/s. With $6\times10^{13}$ cells in the body, we obtain a lower limit $\dot{I}_{lower}$(cell) $> 4\times10^7$ bits/s. When this measure is applied to neurons, we find the dynamic chemical complexity of the brain exceeds by 6 orders of magnitude the complexity we find at the neural level. With the number of neurons in the human brain estimated at $N = 10^{11}$–$10^{13}$, the rate of flux of biochemical reactions in the human brain would be above $4\times10^{18}$–$4\times10^{20}$ bits s$^{-1}$. In a period of 10 years ($\sim 3\times10^8$ sec), this flux of biochemical reactions can produce an amount of information exceeding $I_{biochem}$(brain) $\sim 10^{27}$–$10^{29}$ bits, a much larger quantity than the quantity of information represented in the neural network of the human brain $I_1$(human brain) $\sim 10^{15}$–$10^{17}$ bits obtained above.

In this way, we found a simple solution for resolving the brain–DNA paradox: The genetic complexity of DNA and the algorithmic complexity of the brain are measures that characterize different levels of complexity altogether. Now let us consider the DNA–dynamism paradox.

## 3. The Working Mechanism of the DNA

The question remains: How are we to understand the physical realization of the genome's activity? Actually, how can DNA regulate $4 \times 10^7$ biochemical reactions per second in its host cell? The first problem we need to solve to answer this question is the fact that DNA contains $10^9$ bits information in the sequence of the base pairs, and this is a static form of information. It seems clear that static information cannot elicit any processes. Requiring that DNA control all the biochemical processes of the organism (Woski and Smith, 2002, 28), the DNA must be active, not static. This means that *some part of the DNA must change relative to the others and it is these changes that instruct the biochemical reactions*. In order for these internal changes within DNA to elicit biochemical processes, they must be related to activating processes. The fastest means of such activation are light-induced excitations and electron transfer.

Recently it became clear that long-range single electron transport along the DNA as modulated by intervening sequence and sequence-dependent dynamics might help to switch genes that are far apart on and off (Nunez, Hall and Barton, 1999; Coghlan, 1999). Electron transport and proton translocation are intimately connected with metabolic activity (Demetrius, 2003), and so with the elementary biochemical reactions corresponding to $\dot{I}_{biochem}$. Electronic excited states of complex molecular systems represent the main reservoir of free energy in biologic processes (Korotkov, 2004). Electronic states of complex molecules such as DNA may extend to the whole of the molecule (actually, to the whole of the cell and more); and therefore they are suitable tools to transform sequential static information into a dynamic form. Certainly, not only the static sequential information will play a role, but also the information present in the continuously changing excited states and their biological coupling, which is governed not by the physical laws per se, but by the biological principle. All biochemical reactions can be coupled through electronic states. Our proposal tells that the essence of biological organization is that it couples endergonic to exergonic processes in a suitable manner, preparing the next timestep's input boundary conditions for action by the physical laws by means of these couplings. In this way, DNA becomes able to supply the requirement of timing, determining which chemical reactions should occur in the next timestep. Certainly, DNA cannot do the regulation alone, since its activity must be coupled to cellular organization supplying the necessary chemicals in the necessary places in the right moments, utilizing also a significant part of their thermodynamic capacities. Yet in our model, the dynamic DNA with its active electrons as modulated by sequential information can still maintain its key role of facilitating genetic control over the cellular reactions.

Actually, the timescale for light-induced transfer of electrons (electronic transitions) is $\tau \sim 10^{-12}$ s (Stryer, 1995, 6, Fig. 1–7). The excited electronic states of the cell are modulated by the sequential information of DNA within its collective electronic cloud. This means that biological organization may couple endergonic to exergonic processes that utilize the sequential information of DNA in a way that generates light-induced electronic excitations. The modulated forms of excited electronic states can decay and emit a photon which can activate, e.g., an enzyme, in due time and place. Certainly, the enzymes must be able to act in accordance with instantaneous biological needs. Therefore the whole cell effectively prepares itself as a receptive state in which the activation of a molecule can lead to the realization of the requisite biochemical reaction. Most biochemical reactions of cells are related to enzymes and endergonic–exergonic couplings. This means that the whole cell mobilizes its thermodynamic capacities in accordance with the biological principle.

Now let us estimate the thermodynamic potential of a cell. An average human organism works with an energy flow of $\sim 100$ W distributed on $\sim 10^{14}$ cells. Therefore, an average cell consumes $\sim 10^{-12}$ J s$^{-1}$. This value at $\sim 310°$K corresponds to an energy flow $\dot{E} \sim 3 \times 10^{-14}$ J K$^{-1}$ s$^{-1}$ in entropic units; converting it to units [bits s$^{-1}$] by the simple formula: Information flow in [bits s$^{-1}$] can be obtained from the energy flow in [J K$^{-1}$ s$^{-1}$] when divided by the Boltzmann constant k $\sim 1.38 \times 10^{-23}$ J K$^{-1}$ (see e.g., Brillouin, 1956, pp. 1–3), we obtain that the dynamic thermodynamic capacity of a cell is $\dot{I}_{TD}$ (cell) $\sim 2 \times 10^9$ bits s$^{-1}$. In the foregoing we estimated that more than $4 \times 10^7$ reactions occur per cell per second. It seems to be plausible to estimate that inducing one reaction involves at least one photon and at least one bit of information. Such an estimation yields a lower limit for the biologically utilized information flow of a cell $\dot{I}_{biol}$(cell) > $4 \times 10^7$ bits s$^{-1}$. This estimation shows that cells may utilize a significant part of their thermodynamic potentials for biological aims.

## 4. DNA as the Central Factor of Cells' Cooperation

Regarding the governing activity of DNA, a second question also arises: How does a multicellular organism like a human being *coordinate* the activities of its cells? If the control of the organism is due to DNA, and the individual cell's biochemical reactions are also controlled by the DNA, then we can conjecture that DNA has a twofold function: It has to control *local* cellular processes as well as organizing all these processes into a unique *global* biological organization. This would mean that DNA controls the activity of all the DNA molecules that in turn control the activity of the individual cells. All the DNA molecules are under the control of all the other DNA molecules such that all biochemical reactions serve biological needs useful for the global multicellular organism. The question then becomes: How does the DNA molecule sitting in one cell know about all the chemical reactions occurring in all the other cells governed by the DNA molecules sitting in all the other cells?

It is apparent that multicellular government should be mediated by a coordinating activity between the DNA molecules themselves. Therefore, the DNA molecule sitting in one molecule should follow all the changes occurring in all the other cells. This is a much more demanding task than controlling one cell's activity. This is a task that would require one DNA molecule to act in concert with all the other DNA molecules. This means that instead of $\dot{I}$(DNA) ~ $\dot{I}_{biol}$(cell) ~ $4 \times 10^7$ bits s$^{-1}$, the changes in the internal states of the DNA molecules must be around $\dot{I}$(DNA) ~ $10^{21}$ changes s$^{-1}$.

Thus the question arises: How is it possible to produce $10^{21}$ changes s$^{-1}$ in a DNA molecule if the information content in the sequence of its base pairs is only $10^9$ bits? We point out that this requirement can be fulfilled if the timescale required to induce a change is $10^{-12}$ s. Actually, the DNA molecule is able to realize ~$10^{21}$ changes per second by the fastest known means of biochemical processing — by light-induced electronic excitations (Stryer, 1995, 6, Fig. 1-6). This is what one can expect if these changes are generated by biological organization that is itself the manifestation of the first principle of biology (e.g., the Bauer principle). *The action of first principles does not require mechanisms in order to be achieved.* The first principle of physics, the action principle, does not require a computer built into each elementary particle that would compute which way the particle ought to go. Instead, particles work with the action principle because *the action principle is 'built into' the elementary particles*.

In the hypothetical absence of a first principle of biology, an explanation of the governance of electronic excitation states, and the activation of photons, enzymes, and proteins that correspond to biologically optimal trajectories, would be a computationally unsolvable problem. To govern the states of DNA's $10^9$ base pairs in a way that each changes in every time step of ~$10^{-12}$ s should occur in the biologically optimal manner requires the presence of a first principle of biology in action. There is no way to solve this enormous computational problem except by positing a first principle of biology — the most economic and (we suspect) the only possible solution.

## 5. On the Activity of the First Principles

The activity of the action principle is best understood by the Feynman path-integral interpretation. It states that the action principle has a quantum physical origin: Each elementary particle emits virtual particles which map all possible paths of the whole environment, the collective behavior of which summarizes all possible quantum mechanical paths and realizes the extremum of action having a dimension [energy][time]. The extremum usually indicates the minimum in physical situations (Feynman, Hibbs, 1965, 245). When Feynman introduced the path-integral principle, he pointed out that to be able to follow the principle of least action, quanta must 'virtually' go over all the possible histories, and then these add up to the 'actual' shortest route. The precondition of such an adding up is that in the course of surveying all the possible routes, each quantum virtually travels over all the routes — certainly, at a speed much larger than the velocity of light. Therefore, it is usually said that the Feynman path integral method offers only a model. We note that this model works not only as one of the best of physics, the most exact branch of the natural sciences, but it is the core technology of modern theoretical physics (Moore, 1996; Taylor, 2003; Moore, 2004).

*The point is that the first principles and the faster-than-light virtual actions governing actual interactions are two sides of the same coin.*

The integral form of the action principle contains a non-negligible advantage over its formulation in differential equations. Differential equations need definite initial conditions, while the integral formalism — virtually — includes informative interactions with a large set of the environment. Integral principles are independent from coordinates, and therefore they can cope with time-dependent boundary conditions as well. The apparent teleological behavior of living organisms may correspond to computational processes determined at the organism level, where the organism acts as an agent, following its own interests and biological needs, such as survival. Once the biologically favorable endpoint of a biological process is prescribed by the organism at the organism level, the problem is simplified; and with the help of the action principle of physics it becomes possible to determine the trajectory to be followed, which involves the organism's satisfaction of its biological needs by rearranging its internal physical environment.

Let us consider an extended version of the Galilei experiment. In this example, not only inanimate things, but also living organisms are dropped from the Pisa tower. When a living bird is dropped from a height, it will decide which way to go — i.e., to fall to the ground or fly away — and the biological decision will be realized directly by the action principle since this is the most economic way to organize the physical activities of the bird. Clearly the action principle is subservient to biological organization. The decision of the bird about the optimal endpoint is obtained through switching out into a new context and wider perspectives. If the bird before dropped from a height were in a state of short-period or quasi-instantaneous perspectives, at the moment when dropped it switches into a much wider perspective, the perspective of its life. In this new perspective, it decides about the endpoint of its flight. Once the endpoint is determined, the physical realization of the trajectory best suited to the selected endpoint — survival in this case — can be supplied by the action principle of physics as the most economic solution. This economy extends not only to the minimization of action, but also to the minimization of biological interventions. Absent an integral principle, the bird would need to compute in every time step all the necessary boundary conditions for all the physical processes occurring within it that may lead to the optimal biological solution regarding the selected endpoint of the flight. This task seems to require an overly demanding computational faculty, since every elementary process should result from previous computations, taking into account all the combinatorial possibilities. Without doubt, the action principle is suitable to biological government through endpoint determination. It seems that now it is only a step forward to assume that the action principle is tailored just for biological purposes. Let us consider this hypothesis a bit more closely.

We propose a simple but powerful qualitative idea: that the first principle of biology arises at the other extremum of integrated action [energy][time], which corresponds to the maximum of integrated action instead of its minimum. Biological organization acts to secure the optimal conditions for life. Remarkably, it is the maximum of integrated [energy][time] that corresponds to the optimal quality of life, as is easily seen with the help of an example.

Life's quality is the better, the more the number of years we live and the more free energy we have by which to live our years. Therefore, life's quality can be measured by a quantity: integrated [energy][time]. Now if the first principle of biology acts to realize the maximum of vitality (defined as the distance of the living organism from the deathly thermodynamic equilibrium) not only in the momentary context, but also in the context of our full lifespan, then it appears the Bauer principle likewise would require a maximum of a quantity that also takes the form of integrated action [energy][time].

We note that vitality — the distance of the living organism from thermodynamic equilibrium — is different from Schrödinger's negative entropy, which measures the distance from thermodynamic equilibrium for isolated systems in a static or closed environment. By definition, isolated systems cannot exchange matter and energy with their environment.

We acknowledge a similar idea of Rashevsky (1973, 177), the founder of mathematical biology, who worked out a unified approach to physics, biology, and sociology. In this work one of his main achievements is his Postulate 1 (ibid., 185), which states: "The evolution or time course of change in any organismic set is characterized by the requirement that during the total time of the existence of an organismic set the total number of different relations involved should have a maximum." In comparison, our formulation tells: "*Living organisms invest internal work from their free energy resources against the equilibrium that should be reached on the basis of the given initial state and the physico-chemical laws in order to maximize the total integrated action.*"

We saw that the Bauer principle is a variant of the action principle of physics, since the Bauer principle maximizes the action, while in physics the action is usually minimized. Certainly if DNA works with the Bauer principle, and the Bauer principle is a kind of action principle, then the Bauer principle must act in the same way as the action principle does: by virtual interactions able to map all the universe instantaneously.

The idea that Nature pursues economy in all her workings is one of the oldest principles of theoretical science. In a physical problem, the "action" in the action principle represents a cost. For example, in Fermat's principle the cost expended by a light ray moving along its path is the transit time. In many engineering problems time as well as energy plays the role of a cost, and in such cases the most economic solution is what minimizes the product [energy][time]. This recognition triggered suspicions of the presence of economical aspects.

But it seems a bigger problem is involved here: Such economical aspects are alien to the physicalist world picture and so usually remain undefined and unclear. The action principle has a teleological character that does not fit into the present conceptual scheme of physicalism. Instead of predicting the future from initial conditions, as for instance when working with differential equations, from the more fundamental viewpoint of the action principle, the system starts with a combination of initial conditions and final conditions and the task is to find the path in between them, *as if* the system somehow knows where it wants to go. Actually, it is generally argued that the system does not need the advantage of prior knowledge where it has to go, since the path integral calculates the probability amplitude for any given process. But the real problem is that these probability amplitudes would need (ostensibly) to be calculated for all the possible paths in between.

These problematic and unresolved aspects led to the strange situation in which the highest achievement of physics, its first principle, remained largely ignored in its real impact, and so left without proper interpretation. Additionally, by our proposal action is a basic concept in biology just as it is in physics, since the first principle of biology maximizes it. In this way, an elegant situation arises in which the two known first principles of Nature both take the same quantity — action — to the extreme, each in its own way.

Moreover, since biological organization maximizes action, it is a natural requirement of the Bauer principle that *once the coupling that can lead to the biologically optimal solution is decided and the biological endpoint corresponding to optimal life conditions is determined, the physical realization of decisions should occur with the minimum of action*, since this circumstance is the condition of the requirement that life ever navigates towards its maximum by means of efficient action within the context of the systemic whole. These considerations seem to favor our proposal that the action principle of physics arises as a natural consequence of the Bauer principle.

Returning to the working mechanism of DNA, we can illustrate it with the help of a parable. DNA acts as a watchtower having $10^9$ lamps. Each lamp is switched on or off in every time step, at a frequency interval of roughly $10^{-12}$ s. Collectively the lamps illuminate an enormous biologically useful information flux. Now switching "on or off" are already decisions that one expects would be preceded by information processing. But since the switching on or off occurs at the speed of light, and no physical event can occur at a speed faster than the speed of light, there is no physical possibility for information processing to precede the event characterized by the condition "on or off."

Therefore we realize that DNA works with the help of a factor that is utterly beyond DNA's or any other material system's physical capabilities. On our view, this something is immaterial yet effective and belongs to science — this something we have denoted as a *first principle.* The first principle of biology acts as a deeper intelligence of the "vacuum," in the sense that it virtually maps all the possible histories; summarizes the results of this mapping on its own basis; then decides about the biological endpoint; and from there, "chooses" the optimum physically realizable path.

Let us keep in mind that the action principle *also* acts as a mediated, faster-than-light virtual process that maps the entire universe before the quantum makes its "decision" as to which way to go. Therefore, the strange ability of DNA processing information faster than the speed of light is in good company; namely, in the company of the action principle, the formulation of which by the path-integral formalism has led to some of the greatest achievements of physics. Just as physics resonates to its first principle, the action principle, so DNA acts by the first principle of biology, the life principle (about the life principle see the theoretical biology of Bauer, 1935/1967).

Now if virtual particles correspond to the vacuum, and virtual particles correspond not only to the physical but also to the biological first principle, than *the vacuum must have not only a physical but also a biological nature.* We revealed the existence of the biological vacuum.

Perhaps we are now aware that we have penetrated the realm of ultimate reality. Nevertheless, we feel safe because we find ourselves in the best company, together with the first principles of modern science, which are the safest ground yet achieved by science. Now it is clear that if DNA is governed by the life principle, then the theory of evolution and theoretical physics work with toolkits from partially different conceptual storehouses. We do not require the introduction of new elements into scientific research; we just take into account the most effective tools of science, the first principles, at their real face value. Our solution is the most economic possible: *It extends the action principle just* *by one step so to include the selection of biological endpoints.* This solution offers the most effective way to integrate the action principle and Ervin Bauer's life principle.

## 6. Evolution or Divine Action? Complexity Jumps in the History of Life

As Maynard Smith and Eörs Szathmáry show in their Table 1.1 (1995, 5), the coding part of the bacterial genome has $N_{bp}$(bacterial) ~ $4 \times 10^6$ base pairs and the human genome has $N_{bp}$(human) ~ $6 \times 10^8$ base pairs. We find it remarkable that the size of the coding DNA shows a mere hundredfold increase from bacteria to humans, from $4 \times 10^6$ base pairs to $6 \times 10^8$ base pairs. It is widely thought that terrestrial life was already present within 100 million years after the solidification of the Earth's crust. Now we only point out the obvious, below we will argue for it quantitatively: Evolution is possible only in the presence of life. Chemical abiogenesis cannot produce algorithmic and genetic complexity from morphological complexity, just as emergent phenomena cannot produce the laws of nature.

In this context, it is important to take into account the fundamental fact that the laws of physics have a very low information content, since their algorithmic complexity can be characterized by a computer program less than a thousand characters (Chaitin, 1985). In a personal communication, Chaitin wrote (2004): "My paper on physics was never published, only as an IBM report. In it I took: Newton's laws, Maxwell's laws, the Schrödinger equation, and Einstein's field equations for curved spacetime near a black hole, and solved them numerically, giving 'motion-picture' solutions. The programs, which were written in an obsolete computer programming language APL2 at roughly the level of Mathematica, were all about half a page long, which is amazingly simple."

Now one may estimate the complexity of a page as approximately $2 \times 10^3$ bits, since the average rate of information processing in reading is about 50 bits s$^{-1}$ (Breuer, 1995, 13); and so at a reading rate of 1.5 pages per minute, the information content of a page is about $10^3$ bits. Taking a page from Chaitin, we thus surmise that the algorithmic complexity of physical equations is surprisingly low, $I_{algorithmic}$(physical equations) ~ $10^3$ bits. We think that the low algorithmic complexity of the physical laws is shown also by the fact that present-day physical cosmological models fail to account for such basic phenomena as stellar activity, not to mention protein-based life, with its extremely rich variability.

In contrast to the basic claim of physicalism, the failure of cosmological models with respect to biology is not a practical, but a principal one. Physical models are not only unable to predict biological phenomena, but there is no physical model that can calculate from the positions of particles whether an animal will go left or right from its initial state, nor account for any characteristics of the trajectory of the animal. Moreover, seemingly there are no scientific works in progress in this particular field. No physical models are under construction that could predict such simple phenomena. In contrast, our biological approach is able to work out such a model. Once the main range of the biological endpoint is determined by biological needs and aims, the arising physical trajectory of a bird dropped from a height can be derived (Grandpierre, manuscript).

This suggests that information-producing complexity jumps simply are not possible in the case of physical systems, for such jumps are novel *increases* of complexity: They are ever jumps "up".

Certainly, the observed flow of environmental information is enormous, but it is morphological information. Now since we cannot expect that Big Bang (or recycling) cosmological models obtained initial conditions corresponding to an algorithmic complexity higher than the algorithmic complexity of the physical laws themselves, we can estimate that the complexity measure of physics — initial and boundary conditions and physical equations included — is also about $I(\text{physics}) \sim 10^3$ bits.

The central thesis of physicalism proclaims the causal closure of the physical. Ashby's Law (Ashby, 1962) and Kahre's Law of Diminishing Information (Kahre, 2002) stated that physical systems cannot produce more information at their output than was present at their input. This means that for physical systems, complexity jumps are simply not possible. Therefore the fact that we observe complexity up-jumps here on Earth strongly indicates the presence of life.

The comparison of machines and living organisms can shed light on the nature of biological organization. Once the machine is constructed, its algorithmic complexity is fixed. Even in machines programmed with "learning abilities," only phenomenal data can be involved, and such data cannot increase algorithmic complexity. In contrast, biological organization is able to increase not only algorithmic, but also genetic complexity, as shown by the blossoming of complexity in plants, animals, and in evolution generally.

We point out that there was a much greater complexity up-jump between the early Earth without life and the first bacteria (from $10^3$ bits to $4\times10^6$ bits, a jump of $J_1(10^8$ years$) \sim 4\times10^3$, within about $t_1 \sim 10^8$ years) than between the first bacteria and humans (from $4\times10^6$ bits to $6\times10^8$ bits, a jump of $J_2(4\times10^9$ years$) \sim 150$, during $t_2 \sim 4\times10^9$ years). It means that abiogenetic evolution should be more than thousand-fold faster ($J_1/J_2*t_2/t_1 \sim 1067$) than biological evolution. This fact seems strange, since we recognize that chemical abiogenesis appears (in principle) unable to accelerate the evolution of complexity faster than the evolution of life itself. The question inevitably arises: How can we expect chemical evolution to reach a twenty-seven times higher increase in complexity within a forty times shorter time period than life itself managed to do?

In this context, an example may be enlightening. Hoyle (1983, 243) pointed out that to solve the Rubik cube by one random step in every second, it would take $1.35\times10^{12}$ years. The chance against each move producing perfect color matching for all the cube's faces is about $5\times10^{19}$ to 1. Now if an intelligence is present, reporting after each move whether it was successful or not, reckoning 1 minute for each successful move and, say, 120 moves to reach the solution, the solution of the same Rubik cube may be reached within 2 hours. This fact indicates that the presence of life or intelligence can accelerate evolution in a rate higher than $\sim 6v10^{15}$ ($1.35\times10^{12}$ years/2 hours $\sim 5.9*10^{15}$). Certainly, the abiotic processes are not completely random — modifying the success ratio with and without intelligence from about $10^{16}$ to somewhat lower.

There exists a popular example of monkeys that can type Shakespeare's complete oeuvre on a typewriter. Actually, to type only one sentence from *Hamlet*, consisting of 40 letters, each selected from 30 possibilities, it would be necessary to realize $30^{40} \sim 10^{59}$ trials. Let us assume that we have ten billion monkeys — that is, rather more monkeys than there are currently people in the world. And let us imagine each monkey hits one key per second. Let us further assume that they never stop to sleep or eat or anything else. It will still take more than $10^{49}$ seconds before one of the monkeys has the luck to hit on the right sequence. Now one year is about 32 million seconds, so it will take our world population of monkeys about $3\times10^{41}$ years to get there.

Now how would it be possible that the absence of monkeys and typewriters — corresponding to the case of chemical abiogenesis — could accelerate the process to write an amount of information corresponding to Shakespeare's whole *Hamlet*? Certainly, one cannot expect that chemical evolution would be able to produce useful amounts of genetic complexity in the absence of agents. Even in the presence of "inanimate agents," it seems highly implausible to expect that the accumulation rate of genetic information by chemical abiogenesis in an assumedly *physical* environment (information accumulation in physical systems is excluded by Ashby's Law, Kahre's Law, and causal closure) could produce a meaningfully higher jump in genetic information than the jump produced by life during its $4\times10^9$ years of evolution. Why should "inanimate agents," if they exist at all, be more efficient than living agents possessing much higher genetic complexity?

The complexity measures and their analysis presented here argue that the birth of life here on the Earth is counterindicated by two strong constraints. The first constraint arises from Ashby's Law, Kahre's Law, and the hypothesis of the causal closure of the physical. This constraint tells that *there is no free lunch of algorithmic and deeper complexities*. Algorithmic and deeper (genetic, principal) complexities can be generated only in the presence of life. The second constraint arises from the comparison of the complexity jumps $J_1$ and $J_2$ estimated above. It remains the task of the believers of abiogenesis theory to show how abiogenetic evolution can be more than $10^3$ faster than biological evolution, even if biological evolution is indicated to be faster by a factor $\sim 6\times10^{15}$ or higher (see the example of writing Shakespeare's *Hamlet* by abiogenetic ways, without monkeys and typewriters).

Our results quantitatively argue that life at its fundaments acts by almost fully informed light, filled with information produced in the virtual reality of the quantum vacuum fields, governed by the first principle of biology: the Bauer principle. This aspect lends an intelligent character to the dynamic form of genetic complexity. On the other hand, we also acknowledge the role of natural selection in the evolution of genetic complexity. The arguments presented here together seem to point towards the hypothesis that evolution understood as the increase of complexity is possible *only* in the presence of life.

We found strong quantitative indications showing that life in the virtual quantum vacuum belongs to the utmost foundation of the Universe, and it is this cosmic life that is responsible for the evolution of genetic complexity from bacteria to humans here on Earth.

In the picture outlined above, the natural sciences do not exhaust their possibilities in physics and applied physics, but involve biology (and arguably psychology or the science of self-consciousness) as well. The integral and natural approach presented here differs from the intelligent design theories in that it does not refer to an intelligent factor *beyond* Nature, but involves a deeper intelligence represented *within* Nature *in the form of first principles* from which all the fundamental laws of physics and biology can be deduced and tested by empirical experience.

The integral approach opens new perspectives before modeling biological behavior, collecting, grouping and explaining yet unexplained phenomena. Moreover, it is able to offer predictions on the basis of these first principles, which can then be tested by empirical measurements. For example, the Bauer-principle predicts that in living organisms all the processes governed by physical laws (decay of proteins, heat radiated away, entropy increase, etc.) will be accompanied by processes that will practically compensate the decrease of distance from thermodynamic equilibrium. In living organisms, the fundamental coupling is not between spatiotemporal coordinates, but between global thermodynamic variables of the organisms.

More concretely, the integral approach offers an explanatory model for biological phenomena such as a bird dropped from a height (a prototype of all physical processes). All physical phenomena proceed towards equilibrium: a stone dropped from a height, a warm pond cooling at the onset of evening cold, or a sugar cube dissolving in a cup of tea. All these processes have their characteristic timescales. In living organisms, many similar processes occur, but most of them are compensated for by thermodynamically uphill processes. Therefore, the physical approach cannot model biological behavior at the level of global organization, such as the trajectory of a thirsty animal towards a river. In contrast, the integral model allows the existence of biological aspects such as the determination of biological endpoints by the organism itself. Once the existence of biologically determined endpoints is acknowledged, the integral approach facilitates the determination of the physical trajectory corresponding to the given biological end. The integral model offers the simplest and most effective approach; it is able to predict and it can be tested, therefore fulfilling the sharpest criteria of methodological science. The utmost simplicity at the level of seeming utmost complexity is also explained by the biological principle.

## 7. Summary

We derived quantitative measures of algorithmic complexity of the human brain and of genetic complexity of the human organism. Already these simple complexity measures indicate a paradox of how the brain's complexity can be larger than the complexity of the whole organism. The resolution of this paradox leads us to recognize that genetic information corresponds to a deeper than algorithmic level of complexity. In our consideration of how DNA can regulate the biochemical activities of the organism, we point out that "static" DNA must be complemented by a dynamic complexity that corresponds to its static sequential information content. Numerical estimations show that DNA's changes are regulated by light that is almost fully informed by biologically useful information. Qualitative arguments based on Ashby's Law, Kahre's Law and the hypothesis of causal closure as well as quantitative arguments based on complexity measures of life show strong indications against abiogenesis. We point out that complexity measures show that genetic complexity cannot be produced from environmental effects alone but are governed by the already known, quantitatively formulated first principle of theoretical biology (Bauer, 1935/1967, 51) that is on a similarly firm footing as modern theoretical physics. This means that biological evolution is governed fundamentally by the Bauer principle, and natural selection represents an important, but secondary factor.

Key words: algorithmic complexity, dynamic complexity measures, genetic complexity, working mechanism of DNA.

## 8. Acknowledgements

It is a pleasure to express my thanks to my friend Jean Drew for inspiring communications and correcting the English.

**References**

Ashby, W.R. (1962) in: H.V. Foster, and G.W Zopf (eds.): *Principles of Self-Organization*. Pergamon Press, Oxford, 255.

Bauer, E. (1935/1967) *Theoretical Biology* (1935: in Russian; 1967: in Hungarian) Akadémiai Kiadó, Budapest, 51.

Breuer, H. (1995) *Informatik. DTV-Atlas zur Informatik*. Deutscher Taschenbuch Verlag. Kg, München, 13.

Brillouin, L. (1956) *Science and Information Theory*. Academic Press, New York, 1-3.

Chaitin, G.J. (1985) An APL2 gallery of mathematical physics — a course outline. Proc. Japan '85 APL Symposium, Publ. N:GE18-9948-0 IBM Japan, 1-26.

Coghlan, A. (1999) Electric DNA, New Scientist, 13 Feb 1999.

Demetrius, L. (2003) Quantum statistics and allometric scaling of organisms, Physica A 322, 477-490.

Denbigh, K.G. (1975) *An Inventive Universe*. George Braziller, New York, 99.

Elitzur, A.C. (1994) Let There Be Life. Thermodynamic Reflections on Biogenesis and Evolution. *J. theor. Biol.* 168, 429-459.

Feynman, R.P. and Hibbs, A.R. (1965) *Quantum Mechanics and Path Integrals*. McGraw-Hill, New York.

Grandpierre, A. (2005) *Our Life and the All-pervasive Order* (in Hungarian). Barrus Publishers, Budapest.

Hoyle, F. (1983) *The Intelligent Universe*. 243, see also
http://en.wikipedia.org/wiki/Fred_Hoyle

Kahre, J. (2002) *The Mathematical Theory of Information*. Kluwer Academic Publishers, Boston/Dordrecht/London,
http://www.matheory.info/chapter1.html

Koch, C. and Laurent, G. (1999) Complexity and the Nervous System. Science 284, 96-98.

Kornberg, A. (1989) For the love of enzymes. Harvard University Press, Cambridge, 65.

Korotkov, K. (2004) Assessing Biophysical Energy Transfer Mechanisms in Living Systems: The Basis of Life Processes, Journal of Alternative and Complementary Medicine, 10, 49-57.

Lloyd, S. (2001) Measures of Complexity – a non-exhaustive list, Control Systems Magazine, 21, 7-8,
http://web.mit.edu/esd.83/www/notebook/Complexity.PDF

Maynard Smith, J. and Szatmáry, E. (1995) *The Major Transitions in Evolution.* W. H. Freeman – Spektrum, Oxford, 5.

Maynard Smith, J. (2000) Concept of Information in Biology. Philosophy of Science 67, 177-194.

Moore, T.A. (1996) Least Action Principle, entry in Macmillan Encyclopedia of Physics, John Rigden, ed., Simon and Schuster, Macmillan, Vol. 2, 840.

Moore, T.A. (2004) Getting the most out of the least action: A proposal, Amer. J. Phys. 72, 522-527.

Nunez, M.E., Hall, D.B. and Barton, J.K. (1999) Long-range oxidative damage to DNA: effects of distance and sequence, Chemistry and Biology, 6, 85-97.

Rashevsky, N. (1973) A Unified Approach to Physics, Biology and Sociology, Chapter 2C, in: R. Rosen (ed.) *Foundations of Mathematical Biology*, Vol. III, Academic Press, New York, 177-190.

Shalizi, C.R. (2006) Complexity Measures. http://cscs.umich.edu/~crshalizi/notebooks/complexity-measures.html

Smith, T.E. (1997) *Molecular Cell Biology.* In: T. M. Devlin (ed.) *Textbook of Biochemistry.* 4th edition. Wiley Liss, New York, Ch. 22, p. 921.

Stripling, J. (2004) Introduction to Cognitive Science. Neuroscience, http://comp.uark.edu/~jstripli/CogSci-JS-L1-web.pdf

Stryer, L. (1995) *Biochemistry.* Fourth ed. W. H. Freeman and Co., New York, p. 6 & Fig. 1, 7.

Taylor, E.F. (2003) A call to action. Guest editorial. Amer. J. Phys. 71, 423-425.

Woski, S.A. and Smith, F.J. (2002) *DNA and RNA: Composition and Structure.* in: T. M. Devlin (ed.) *Textbook of Biochemistry.* 5[th] edition. Wiley-Liss, 28.

## *Dialogue*

*Philip Ball*

Understanding life processes in terms of flows of information seems like a fruitful way to proceed in attempting to answer Erwin Schrödinger's question 'what is life?' But it seems to me that this issue is a subtle one that can be obscured as much as it is elucidated by the contemporary emphasis on DNA as a depository of digital information. The relationship between the genetic sequences in DNA and the molecular processes of life is by no means obvious. Take, for example, the way in which cell behaviour is regulated by the operation of protein ion channels. There is certainly a form of logic to this function: the output, transmembrane potential say, can be regulated by a variety of input signals, such as mechanical forces, temperature, the presence of other ions or ligands. The way this transduction occurs is mediated by physical laws of an analog nature, for example Fickian diffusion, thermal fluctuations, electrical gradients. One might argue that, in defining the shape and structure of the ion channel, genetic information in DNA 'governs' the process — but what does that really mean? Does the static information in the gene encoding the protein somehow dictate the temporal switching behaviour of the gated channel over time? No, surely here DNA is more like the medieval God who lays down the initial laws of the universe before sitting back and simply letting them unfold, without constant intervention. The process unfolds because of the nature of the changing environment, acting on a set of preconditions, and not because of tampering by the agency of those preconditions.

This is why I am somewhat puzzled by Grandpierre's conception of DNA 'controlling' cellular biochemical reactions. It rather sounds as though he requires this control to be constantly 'active': a reaction cannot proceed as it should unless DNA is *doing something* to ensure that, which then seems to demand ultrafast processes involving, say, electron transport. This, perhaps, is where we are led by modern biology's insistence on the primacy of DNA as the 'author' of life, so that things cannot be trusted to unfold of their own accord (which is to say, on the basis of simple physicochemical processes). The problem seems to become even more profound once we consider how individual cells coordinate their activity: as Grandpierre asks, "How does the DNA molecule sitting in one cell know about all the chemical reactions occurring in all the other cells?" It seems to me that the answer is that the DNA does not have to 'know' anything; there are mechanisms for cell-cell communication (some of which, in higher organisms, use the very fast transmission of electrochemical potentials to connect remote regions) which have become adapted in such a way as to permit spontaneous, functionally directed self-organization of the multicellular body. One can see primitive self-organized behaviours of this sort in single-celled organisms that can display some collective behaviour when the conditions dictate, such as the slime mold *Dictyostelium discoideum*. The DNA does not have to act as some kind of godlike molecular overseer in these situations.

With this in mind, it seems no longer obvious why one need invoke the kind of biological first principle that Grandpierre discusses towards the end of his article — one that acts as a kind of faster-than-light 'deeper intelligence of the vacuum', deciding about biological endpoints and then mapping the pathway there. (Incidentally, where does the teleological 'biological endpoint' come from? Since when did we need to invoke any prescience to biology in order that it 'works'? Biology is, from moment to moment, surely quite blind, and it is only evolution that has installed an apparent 'purpose' to it all.)

This insistence on absolute and rigid genetic control of biological processes seems to be what motivates the notion of DNA as a 'watchtower' switching on each of its lamps every $10^{-12}$ s. To me, this conjures up the image of an over-zealous lighthouse keeper convinced that no sailor can by themselves work out how to navigate the rocks, whatever the weather or the time of day. At any event, why does this switching have to happen 'at the speed of light', demanding some kind of mysterious superluminal agency that dictates the pattern of switching? I don't understand that. Yet it leads to the even stranger notion of a 'biological vacuum', which seems to me to be some agency that gives organisms biological foresight of the kind that evolution ensures they don't actually need.

Another claim for which I can't find sufficient motivation is the idea that biological systems maximize action. This seems to depend on an assumption that organisms 'try' to live for as long as possible. But that isn't so. They live for as long as is evolutionarily convenient. In conventional Neodarwinian terms, the sole imperative is to maximize the prospects of propagating one's genes: survival is generally a concomitant of this, but not survival at any cost — for example, an organism might stand to benefit more from devoting its limited resources to reproduction than to cell repair. In any event, the hypothesis of maximal action seems to be one that is asserted but never proved.

At root, I am perhaps most perplexed by the notion that algorithmic complexity has to be high to account for biological phenomena. Has it not been one of the underpinnings of complexity science that complex behaviours can arise from simple rules? Grandpierre asserts that no physical models can account for the trajectories of biological organisms. But they can! Models of ant motion, driven by simple ideas such as chemotaxis and random searching, can reproduce the behaviour of ant colonies rather well. For simple organisms such as bacteria, it seems even possible in principle that one might measure from moment to moment all the environmental influences acting on a single cell, and thereby predict its motion with great precision. Certainly, it is not clear why there need be anything mysterious or aphysical about this behaviour.

Grandpierre argues that abiogenesis cannot seem to create, in a sufficiently short time, the complexity we see in life: if I understand correctly, he implies that only life (or 'intelligence') can beget life. To my mind, there are two shortcomings with this. First, it assumes that accumulation of complexity is linear, whereas it now seems that many complex systems possess thresholds above which entirely new modes of behaviour — new capabilities — appear. Secondly, I see no explicit role here for evolution: for the quite remarkable efficiency of searching in the landscape of possibilities for effective 'answers' that is permitted by the rather simple algorithm of random mutation and replication in the face of limited resources. Diversification and complexification are, in this respect, boosted by the fact that every evolutionary step broadens and modifies the landscape in which subsequent steps are taken: evolution does not simply have to respond to a preordained landscape, but to itself. To my mind, "intelligence in Nature" here becomes another God of the gaps, an expression for what we do not yet understand (and what therefore astounds us) about the capacity of the physical world to generate richness and complexity.

*Attila Grandpierre*
I agree with Ball's note that the recent emphasis on DNA as a depository of digital information might be an overstatement. Indeed, as I tried to indicate it in my chapter, perhaps not consequently enough, it is the cell as a whole, with all its constituents and biological couplings, which governs the cell's behavior, and not the DNA alone. For example, I argued that the cell utilizes a significant part of its thermodynamic potential for biological organization. Regarding the problem I raised in my chapter, namely, that the static information of the DNA in itself is not suitable to govern (or participate in) the time sequence of biochemical reactions. In my point of view, this is a fundamental unsolved problem of modern biology. It seems that Ball approaches only the physical aspect of the cell's behavior. Indeed, enlisting the physically influential parameters of the input and the output of the process regulating the behavior of the cell, he implicitly ignores the biological aspects of the problem. The biological aspects of the cell's behavior are related to thermodynamically uphill reactions made possible by biological couplings between endergonic and exergonic reactions. My point is that the DNA also contributes to the biological coupling processes through spontaneous photon emissions and absorptions, electron transfer and many other ways, in coherence with all the biochemical processes, all of which are governed ultimately by an autonomous biological principle. (continued below).

*Philip Ball*
This seems possible, but is there any evidence for it? I'm aware only of, e.g. electron transfer in DNA perhaps playing a role in DNA damage.

*Attila Grandpierre*
It has been pointed out that the genetic code is useless without the supporting cellular machinery — first of all, without properly functioning proteins, required for DNA/RNA functioning (Ben Jacob, Shapira, Tauber, 2006, Seeking the foundations of cognition in bacteria: from Schrödinger's negative entropy to latent information. *Physica A* 359, pp. 495-524; p. 515). It is clear that static, sequential information is useless in generating the dynamic biological organization (Grandpierre, 2007, *NeuroQuantology* 5, pp. 346-362). As I wrote in my chapter: "Recently it became clear that long-range single electron transport along the DNA as modulated by intervening sequence and sequence-dependent dynamics might help to switch genes that are far apart on and off (Nunez, Hall, and Barton, 1999; Coghlan, 1999). Electron transport and proton translocation are intimately connected with metabolic activity (Demetrius, 2003), and so with the elementary biochemical reaction flux".

*Attila Grandpierre* (continued)
This means that, in contrast of Ball's opinion, DNA does not act like the medieval (deistic) God who lays down only the initial laws of the universe before sitting back and simply letting them unfold. Instead, DNA, together with all the subsystems of the cell, continuously changes and these changes add up to the changes of the cell's behavior (continued below).

*Philip Ball*
I still don't see why this is necessary. Isn't it one of the basic principles of self-organized systems that they do not need a constant 'hand on the tiller'?

*Attila Grandpierre*
The difference between physical self-organization and biological organization is that physical self-organization does not need a continuous control. In contrast to physical self-organization, in biological organization a continuous flux of information is required to govern biochemical reactions.

*Philip Ball*
To take a simple example, where is the 'information' that promotes lipid assembly?

*Attila Grandpierre*
Lipid assembly corresponds to a concrete, fix form of information. In contrast, another type corresponding of biochemical reactions also exists representing continuously changing information corresponding to continuously changing external and internal conditions. Biological organization governs the relation between such prefixed cycles.

*Philip Ball*
It is not obviously 'in' the genes that encode lipid-synthesis enzymes, but follows from physicochemical principles.

*Attila Grandpierre*
It seems that the example of lipid assembly within given physical conditions simplifies biology to physics by throwing out the baby with the bath water. Biology is present not within the framework of an already definite physical problem, but, on the contrary, biology prepares the conditions for the physical laws to act. Biology is the control science of physics. The point is that biology starts 'before' physics, preparing the input conditions for the physical laws. Once the physical conditions are suitably prepared by biological organization, the rest is physics, I agree. I am speaking about the biological aspect, while you seem to be concerned with the physical aspect of the problem. It seems we are speaking about different aspects of the same subject.

*Attila Grandpierre* (continued)
I emphasize that not only the outer environment is influential in determining the behavior of the cell, but the internal environment, too, the determination is not absolute, and it does not occur only on the basis of physical laws. Physical laws are the ultimate, instantaneous tools of biological reactions, but the conditions within which the physical laws act are governed by biological couplings in a way that within the continuously changing biological conditions the physical laws result a biological behavior which is at variance with the physical behavior which would arise in the absence of biological couplings. Cells can utilize the significant part of their thermodynamic potential only through couplings of endergonic and exergonic reactions making it possible to compensate the otherwise inevitable approach towards thermodynamic equilibrium due to entropy increasing physical processes by thermodynamically uphill processes like active transport, regeneration of gradients etc.

*Attila Grandpierre*
In the second paragraph, Ball seems to ignore my point that the whole cell is involved in biological organization when speaking about "Grandpierre's conception of DNA 'controlling' cellular biochemical reactions". My conception is much more modest. I argue only that the DNA must contribute to the government of biological processes.

*Philip Ball*
It rather sounds as though he requires this control to be constantly 'active': a reaction cannot proceed as it should unless DNA is *doing something* to ensure that, which then seems to demand ultrafast processes involving, say, electron transport.

*Attila Grandpierre*
Again, the root of the misunderstanding lies in the different approaches. Ball seems to be involved in the physical approach, considering the moment to moment changes of a certain reaction, with all its physical input conditions already prepared. In contrast, I consider biological behavior in a longer timescale, in the biological context, in which these input conditions are influenced by biological couplings related not only to physical laws and physical conditions but to biological needs and ends as well. (continued below)

> *Philip Ball*
> But is this 'broader picture' not an emergent property?

*Attila Grandpierre*
A property can emerge, but a law cannot. Phenomenal complexity can emerge in a physical process, but algorithmic complexity cannot. Emergence is a process at the morphological/phenomenal level, while laws exist at the level of algorithmic complexity. Physical laws cannot emerge from material properties (like mass, charge, or size). If biological behavior is governed by the Bauer principle, which cannot be derived from the physical principle of the least action, it cannot emerge from physics.

*Philip Ball*
Shaped by intermolecular interactions but not obviously derivable from them?

*Attila Grandpierre*
Recently, it has become clear that simple bacteria can exhibit rich behavior, have internal degrees of freedom, informational capabilities, and freedom to respond by altering itself and others via emission of signals in a self-regulated manner (Ben-Jacob, E. 2003, Bacterial self-organization: co-enhancement of complexification and adaptability in a dynamic environment. *Phil. Trans. R. Soc. Lond. A*, 361:1283-1312). Each bacterium is, by itself, a biotic *autonomous* system, having a certain freedom to select its response to the biochemical messages it receives, including self-alteration, self-plasticity, and *decision making*, permitting *purposeful* alteration of its behavior (Ben Jacob, Aharonov and Shapira, 2005, Bacteria harnessing complexity. *Biofilms*, 1, 239-263; http://star.tau.ac.il/~eshel/papers/11.11.04.pdf). Bacteria are able to reverse the spontaneous course of entropy increase and convert high-entropy inorganic substances into low-entropy life-sustaining molecules (*ibid*.). Similarly, di Primio, Müller, and Lengeler (2000, *SAB2000 Proceedings Supplement, International Society for Adaptive Behavior*, 3-12, http://www.ais.fraunhofer.de/~diprimio/publications/diprimio_MinCog.pdf) have demonstrated that bacteria and other unicellular organisms are *autonomous* and social beings showing cognition in the forms of association, remembering, forgetting, learning, etc., activities that are found in all living organisms. It is widely recognized recently that biochemical reactions are regulated by complex conditions involving practically the whole cell, governed possibly by a yet unknown principle (see Ben Jacob, Shapira and Tauber, 2006, Seeking the foundations of cognition in bacteria:

from Schrödinger's negative entropy to latent information. *Physica A* 359, 495–524.).

*Philip Ball*
To my mind, DNA simply encodes the local rules that enable such large-scale properties to emerge from the biochemical network. In this sense, DNA does not really contain a 'blueprint' of the organism.

*Attila Grandpierre*
One of the main points in my chapter is that the changes of the DNA that contribute to selecting, timing and localizing biochemical reactions are regulated by the Bauer principle. I agree with Ball's claim that that DNA's instructions are the result of the biochemical network, but only with the reservation that the biological 'network' is in the actual cell an astronomically enormous dynamic flux of biochemical reactions, estimated in my chapter to represent an information flux around $10^7$ bits $s^{-1}$ cell$^{-1}$, which is governed by the Bauer principle. In this respect, the term 'blueprint of the organism' is misleading since suggesting that the biochemical processes are determined by a static material structure similar to a 'blueprint'. Instead, my argument tells that the dynamic flux of biochemical reactions is ultimately governed by the Bauer principle; similarly to physics, since physical behavior is governed by the least action principle. Actually, the 'local rules of the DNA' can be at work only when relying on the universal biological principle; therefore, they are based on the Bauer principle.

(continued)
*Philip Ball*
It seems to me that the answer is that the DNA does not have to 'know' anything; there are mechanisms for cell-cell communication (some of which, in higher organisms, use the very fast transmission of electrochemical potentials to connect remote regions) which have become adapted in such a way as to permit spontaneous, functionally directed self-organization of the multicellular body.

*Attila Grandpierre*
In contrast, I emphasize that DNA has to be informed about the cellular processes, otherwise it cannot contribute to the biological organization in a biologically useful manner. I suggest that ultimately, the DNA is informed by virtual interactions (continued below).

> *Philip Ball*
> What are 'virtual interactions'?
>
> *Attila Grandpierre*
> Virtual interactions are interactions mediated by virtual particles. In the double slit experiment of quantum physics, the action principle is realized by virtual interactions mapping the whole situation and integrating the quantum amplitudes corresponding to all possible trajectories of the quanta. Feynman's path integral approach indicates that quanta explore all possible paths between the initial and end states (Taylor, 2003, *Amer. J. Phys.* 2003; 71: 423–425.; Moore, 2004, *Amer. J. Phys.* 2004, 72 : 522–527.), and the resulting path is the integrated sum of all these paths. Virtual interactions are governed in physics by the action principle (Feynman and Hibbs, 1965, *Quantum Mechanics and Path Integrals*, McGraw-Hill.

(continued) that determine the biological couplings, which govern biological processes like cell-to-cell communication etc. It seems that the conflict between the physical and the biological approach is manifest when Ball speaks about permitting "spontaneous, functionally directed self-organization of the multicellular body". The point is that physical self-organization is a process which is governed by physical laws. In contrast, biological or functional organization serves a biological need or end, like a biological function, a concept alien to physics. Therefore functionally directed self-organization is not a physical process. It is possible only with the assistance of biological couplings which most fundamentally determine the input conditions for the physical laws (continued below).

> *Philip Ball*
> Aha! Maybe this is really where our views diverge. It seems to me that all one needs to obtain function is physical self-organization coupled to selective pressure.
>
> *Attila Grandpierre*
> As I pointed out above, physical self-organization occurs only occasionally, while life requires continuous modifications of the input conditions of physical laws. Function is clearly a biological concept. In general, I define biological function as consisting from processes solving a biologically useful task. In cases of physical self-organization like formation of snowflake patterns, or Benard convection cells, we cannot speak about biological functions. Certainly, we cannot speak about 'selection pressure' in case of physical self-organization processes. Therefore, it is possible to couple physical self-organization to selective pressure only in case of living organisms. Now living organisms are governed by the Bauer principle, and so it is not allowed to exclude the biological principle from the picture by substituting it with physical self-organization plus selective processes.

*Philip Ball*
Certainly, that seems to work in *in vitro* chemical evolution.

*Attila Grandpierre*
Certainly, one cannot speak about biological functions in case of chemical evolution. I note that this point requires ramifications in the context of the problem of continuity of life with the apparently inanimate world, as many scientists suggested; e.g. Editorial, 2007, The meaning of 'life', *Nature* 447, pp. 1031-1032.

(continued)
*Philip Ball*
"With this in mind, it seems no longer obvious why one need invoke the kind of biological first principle that Grandpierre discusses towards the end of his article,…deciding about biological endpoints and then mapping the pathway there".

*Attila Grandpierre*
Certainly, with this in mind, i.e. with the physical approach in mind and overlooking the fundamental difference between the physical and biological self-organization, it seems no longer obvious the need for an autonomous biological organization.

*Philip Ball*
Incidentally, where does the teleological 'biological endpoint' come from? Since when did we need to invoke any prescience to biology in order that it 'works'?

*Attila Grandpierre*
Yes, the teleological biological endpoint comes from the biological principle, the most action principle, as well as from the autonomous selection from the range of all possible biologically prescribed biological endpoints by the organism. For example, when a bird is dropped from a height from the Pisa tower, in the first moment it moves exactly like a dead bird or a stone, i.e. following the law of free fall. But as moment comes after moment, slowly the bird starts to modify its internal structure with the help of biological couplings, and with the help of the energy resources arising from the exergonic reactions make it possible to initiate biologically useful endergonic reactions. With the help of self-initiated internal changes, modifying its external shape, the bird extends its wings and modifies its trajectory. In contrast to widespread opinions, teleology is not alien in science, see e.g. Thomas Nagel (1979, *Teleology Revisited and Other Essays in the Philosophy and History of Science.* New York, Columbia University Press, p. 278.). Teleology is directly related to functions (continued below).

*Philip Ball*
But evolutionary theory surely shows that function does not imply teleology?

*Attila Grandpierre*
Evolutionary theory does not explain the origin and nature of biological functions. It only indicates that systems of biological functions, once they exist, can evolve. Machines receive functions only by human activity. Machine's functions represent human teleology, or purpose. Biological functions serve biological purposes (see also Buller, D. J. 2002, Function and Teleology, in: *Encyclopedia of Life Sciences*, p. 393), since ultimately they represent biological endpoint selection as input for the action principle. Without biological endpoint selection serving the most action principle, only the least action principle would be at work, and so only physical processes could occur, driving the organism towards thermodynamic equilibrium and death. Therefore, biological functions inevitably represent biological teleology.

(continued)
*Attila Grandpierre*
…and biology is the science of mechanisms and functions. "Biology, the scientific study of living organisms, is concerned with both mechanistic explanations and with the study of function." (Purves, Orians and Heller, 1992, Life. The Science of Biology, 1) Therefore, it would be unscientific to ignore the directly teleological aspects of living organisms. The characteristic difference between the physical approach and the biological one is present again in the sentence of Ball:

*Philip Ball*
"Biology is, from moment to moment, surely quite blind".

*Attila Grandpierre*
In physics, the differential equations plus completely blind random fluctuations govern the behavior from moment to moment. In biology, the case is very similar for changes occurring from moment to moment, but very different for changes occurring from the initial state to the final state. Living organisms are able to mobilize their internal energy resources, and, through biological couplings, they can initiate investment of biologically useful work (to recharge their potentials, or to fly away from the free fall trajectory). While the flight of the fallen bird can be regarded from moment to moment as determined by physical conditions, biological couplings act in order to modify the input conditions of the physical equations continuously and systematically, deteriorating the bird from the physical trajectory and making it able to follow any trajectory consistent with the biological principle. (continued below).

*Philip Ball*
To my mind, the bird has been given a 'flying instinct' by evolution, which it exercises. This 'flying instinct' is a consequence of blind exploration of evolutionary space, coupled to selection.

*Attila Grandpierre*
In the first chapter "Gaps and Inconsistencies in Modern Evolutionary Thought" of the 728 pages book of West-Eberhard (2003, *Developmental Plasticity and Evolution*, Oxford University Press) a whole list of basic problems of evolutionary theory are presented. It becomes more and more clear that Darwinian theory is so logically flabby it can "explain" anything by subtly changing the terms of the debate. Evolutionary theory can show only that systems of functions may evolve in a changing environment, but does not explain how an individual cell selects from the astronomically large domain of biological possibilities. Evolutionary theory concerns only the historical life forms appearing on the earth. It considers only a part of biological phenomena, instead of working out the general theory of biological processes and deriving the more special phenomena from the more general laws as it is possible in physics. In contrast, the theoretical biology of Ervin Bauer established the most universal law of biology in an exact manner which is quite compatible with the exactness of physics. These arguments indicate that selection is not the cause but the result of biological organization. Therefore, ultimately, the flying instinct, together with the phenomenon of evolution, is based on the Bauer principle.

*Philip Ball*
I think the example is clearer if we take a less emotive case: a bacterium released near a source of nutrient, to which it can find its way by chemotaxis.

*Attila Grandpierre*
I repeat my answer given in the previous round: chemotaxis is already a biological input.

*Philip Ball*
One can design a robot, or most probably soon an artificial cell, to do that. It is all purely mechanical.

*Attila Grandpierre*
No, it is not mechanical, even not physical; ultimately, it is always the decision of the organism to decide to go for the food. Deciding to go for the food is prescribed not in a coercive manner like the laws of physics but depend on the autonomy of the organism, too. Biological decisions correspond to the most action principle. Once the autonomous organism decided to go for the food, the rest is physics, because the endpoint is decided, and so the most action principle requires the work on the basis of the least action principle in order to secure the most action possible for the organism. Therefore, the first principle of physics is derivable from the most action principle of biology.

*Philip Ball*
In nature, it's surely now clear why the 'watchmaker' involved in 'design' can be blind.

*Attila Grandpierre*
I repeat: physics comes only after the biological endpoint is determined. One can close the eye only if it already selected what to do.

(continued)

*Philip Ball*
"It is only evolution that has installed an apparent 'purpose' to it all."

*Attila Grandpierre*
Evolution acts only on the species and not on the individual. The trajectory of a fallen bird is not prescribed by evolution in all its details. Evolution cannot play the role of an absolute and rigid control of biological processes.

*Philip Ball*
Evolution nevertheless shaped all individuals in a species. And I'm not sure it's true to say that evolution can't explain the bird's trajectory.

*Attila Grandpierre*
Definitely, evolution acts only at the level of species, and cannot determine the trajectory of a concrete bird in all its details. Even in the same environment, the bird dropped from a height can fly to many directions and can select many different trajectories.

*Philip Ball*
Evolution can explain how the bird gets the instincts and apparatus for flying; the rest is mechanics.

*Attila Grandpierre*
This claim can be stated only when ignoring the most fundamental aspect of biological organization, namely, biological endpoint selection.

Philip Ball
Yet it leads to the even stranger notion of a 'biological vacuum', which seems to me to be some agency that gives organisms biological foresight of the kind that evolution ensures they don't actually need.

*Attila Grandpierre*
Evolution is a special aspect of biology. The theory of evolution has quite a different character from theoretical physics. The fundamental laws of physics are the most general laws of physical phenomena, while the theory of evolution considers only a special phenomenon, the evolution of species, and only within special conditions present in the Earth, and so it fundamentally lacks due generality. Biology will reach the position of a mature natural science only if it finds the most general laws of biological behavior. Once theoretical biology will develop, finding its first principle, it will be a more fundamental theory than the theory of evolution. Therefore, rejecting the possibility to consider the first principle of biology on the basis of sticking to a special theory has a corollary hindering the development of natural science (continued below).

*Philip Ball*
I'm not sure I agree that evolutionary theory need be seen as a 'special case' in biology, or that it precludes other views on biological behaviour, such as a systems view of the cell or explanations for group coordination.

*Attila Grandpierre*
The fundamental equations of physics determine the most general laws of all possible physical objects. In contrast, evolutionary theory is not the theory of the most general laws of motion of all living organisms possible. It is only the theory of evolution of species of the biosphere present on the Earth. As it is not correct to claim that the semiempirical theory of the growth of the tree is the ultimate theory of the tree, it is also not correct to claim that the semiempirical theory of evolution is the ultimate theory of biology.

(continued)
*Philip Ball*
Another claim for which I can't find sufficient motivation is the idea that biological systems maximize action. This seems to depend on an assumption that organisms 'try' to live for as long as possible. But that isn't so. They live for as long as is evolutionarily convenient.

*Attila Grandpierre*
Again, the subject of evolution is the species, and not the individual. In most cases it is an irresistible biological urge for the individual to survive, independently of its "evolutionary convenient" age. Even if *Homo sapiens* would have an evolutionary convenient life span, let us take 80 years, most older people still want to survive. (continued).

*Philip Ball*
Ultimately the average lifespan is determined by evolutionary factors, regardless of the fact that we have a survival instinct.

*Attila Grandpierre*
Again, we have to distinguish between the level of biosphere, the level of a certain species and the level of a certain individual. The average lifespan of a species is determined at the level of the biosphere, which can be regarded also as a living organism; see Shapiro, R. 1998, *Planetary Dreams*. Therefore, the average lifetime of a species corresponds to the most action principle applied to the biosphere, to the longest lifespan of the biosphere. The same is true on the level of the individual, which also tries to

> live in accordance with the most action principle applied to the individual. Apparently, Ball denies the aim of the individual to live as long and as healthy as possible on the basis of a consideration switching implicitly to a different context, from the individual to the species and the relations of species within the biosphere.
>
> *Philip Ball*
> Of course, humans might one day be able to subvert that technologically (probably to our cost), but we're a special case!
>
> *Attila Grandpierre*
> I hope it is more clear now that I mean not a technological but a biological wish of maximizing survival and health, at the level of the individual.

(continued)
Moreover, the aim of survival only one of the essential aspects maximizing action, since the other factor is to reach the highest amount of available free energy (action=energy*time). Indeed, most living organisms strive not only for survival but for the healthy state in which the available free energy is maximal (continued below).

> *Philip Ball*
> Sounds rather like both stem from the same survival instinct.
>
> *Attila Grandpierre*
> Not necessarily. There are two variables: maximal distance from equilibrium (E) and maximal lifespan (T). Action is roughly the product E*T (more precisely, $\int E(t)dt$). Moreover, my chapter presents not a qualitative argument but a quantitative one formulated mathematically. This can be regarded as a definite achievement.

(continued)
Moreover, Ball considers only extreme cases in which the species and the individual are in conflict (continued below).

> *Philip Ball*
> In any event, the hypothesis of maximal action seems to be one that is asserted but never proved.

*Attila Grandpierre*
As I stated in my chapter, Ervin Bauer (1920, 1935/1967; refs. in the chapter) already established the first principle of biology, proving that all the fundamental life phenomena like metabolism, growth, reproduction, regeneration and death can be derived from it. He also formulated the first principle of biology in a quantitative mathematical form, and made it suitable for calculations of biological processes. Moreover, my re-formulation of the Bauer principle in the form of most action principle is proven in Grandpierre, 2007, *NeuroQuantology* 5, 346-362.

*Philip Ball*
I confess that I'm not familiar with these works – I will need to take a look.

(continued)
*Philip Ball*
At root, I am perhaps most perplexed by the notion that algorithmic complexity has to be high to account for biological phenomena. Has it not been one of the underpinnings of complexity science that complex behaviours can arise from simple rules?

*Attila Grandpierre*
This is a very good question, related to a main aspect of my chapter. Indeed, today it is a dominant view that complexity can arise from simple rules. But one must distinguish between the above weak form of such a statement and a stronger form claiming that all complexities found in nature must be derived from simple rules regulating only the physical properties of the systems and organisms. As I enlightened it in my chapter, algorithmic complexity has a fundamentally different nature from the morphological or phenomenological complexity. Let us take an example. The circle has a low algorithmic complexity (cca. 100 bits), and an extremely high morphological complexity (infinite points, infinite bits). Therefore, it is apparent that a small algorithmic complexity is able to produce an extremely high amount of morphological complexity. In this sense, algorithmic complexity is more fundamental than morphological one. Does it follow from the fact that the circle has a low algorithmic complexity that we must think that all the mathematical functions can be derived from simple rules? No, because, for example, there are many mathematical objects that cannot be given in algebraically closed form. Let us take another example. There are simple machines like a watch having a low algorithmic complexity. Does it follow that we must accept that all machines must have low algorithmic complexity? No, because a computer with higher algorithmic complexity can solve more tasks and more easily than a smaller computer. Moreover, once the machine is ready, its functions are specified, its algorithmic complexity is given. But there are tasks for living organisms requiring revealing a problem, to realize the existence of an unexpected task. Living organisms must continuously solve new and new problems, and problem solving is by definition corresponds to the production of algorithmic complexity. Production of algorithmic complexity is possible only if a still deeper level of complexity (generative complexity) exists which can produce algorithmic complexity on the basis of a unified context corresponding to the generative principle. One of the main points of my chapter is to show that we must realize that algorithmic complexity and generative complexity can be regarded as full members of the conceptual framework of science and they are fundamental aspects of nature. My answer to Ball's problem is that we have to consider systems and organisms with high algorithmic complexity. The algorithmic complexity of a circle or a fractal is low, a watch has a higher algorithmic complexity, a computer still higher, and a living organism still much higher. Indeed, the algorithmic complexity of a living organism must be high to account for biological phenomena (continued below).

> *Philip Ball*
> I need to give this more thought! All I'd say now is that our point of reference in such questions should in my view be a bacterium rather than any higher organism: to my mind, once you have the former, the latter follow…
>
> *Attila Grandpierre*
> There are more and more evidence indicating the existence of autonomous bacterial purpose and intelligence, like e.g. Mathieu and Sonea 1996, Time to drastically change the century-old concept about bacteria, *Science Tribune*, August 1996; Ben-Jacob, 2003, Bacteria harnessing complexity. *Phil. Trans. R. Soc. Lond. A*, 361, pp. 1283-1312; Ben Jacob, Aharonov and Shapira, 2005, Bacterial self-organization: co-enhancement of complexification and adaptability in a dynamic environment. *Biofilms*, 1, 239-263; http://star.tau.ac.il/~eshel/papers/11.11.04.pdf; di Primio, Müller, and Lengeler, 2000, Minimal Cognition in Unicellular Organism. *SAB2000 Proceedings Supplement, International Society for Adaptive Behavior*, 3-12, http://www.ais.fraunhofer.de/~diprimio/publications/diprimio_MinCog.pdf; Ben Jacob, E., Shapira, Y. and Tauber, A. I. 2006, Seeking the foundations of cognition in bacteria: from Schrödinger's negative entropy to latent information. *Physica A* 359, 495–524; Shapiro, J. A. 2007, Bacteria are small but not stupid: cognition, natural genetic engineering and socio-bacteriology. *Stud. Hist. Phil. Biol. & Biomed. Sci.* 38, 807-809; B.

J. Ford, 2004, Are cells ingenious? *Microscope* 52, 135-144; B. J. Ford, 2006, Revealing the ingenuity of the living cell. *Biologist* 53, 221-224.

(continued)
*Philip Ball*
Simple models of ant motion, driven by simple ideas such as chemotaxis and random searching, can reproduce the behaviour of ant colonies rather well.

*Attila Grandpierre*
Again, it seems that Ball thinks in a physical approach without realizing the difference represented in biological behavior. Definitely, when introducing biological concepts like chemotaxis, it is already possible at least in some simple cases to reproduce the observable behavior of ant colonies (but not so well of each individual ant, I guess). My biological approach intends to establish the idea that it is necessary to add suitable biological concepts to our physical vocabulary when attempting to determine biological behavior (see Grandpierre, 2007, *NeuroQuantology* 5, 346-362.).

*Philip Ball*
For simple organisms such as bacteria, it seems even possible in principle that one might sense from moment to moment all the environmental influences acting on a single cell, and thereby predict its motion with great precision.

*Attila Grandpierre*
Again, the difference arises in the approaches. Ball seems to stick to the physical approach corresponding to extremely short timescale changes "from moment to moment". In the long run, in a long enough timescale, it is not possible to determine the behavior of a fallen bird in the absence of suitable biological endpoints (like chemotaxis, reaching the food, escaping from a danger etc.). Therefore, even if it would be possible to predict the behavior of a cell from moment to moment (like the position of a fallen bird in the next instant) approximately, it is not possible to determine the biological behavior of the cell on a purely physical basis on a biological timescale (the trajectory of a fallen bird deteriorating itself from the free fall path) (continued below).

> *Philip Ball*
> Hmm… I'm not sure I agree with the way the problem is posed, as though a complex biological entity like a bird can be compared with an inanimate particle.

*Attila Grandpierre*
The problem is posed in the framework of the Galileo experiment, considering the fall of different objects from the Pisa tower and watching their behavior. The behavior of the bird is compared not to that of an inanimate particle but of an inanimate thing. The problem is posed in accordance to the central task of science, namely, finding the laws of nature governing observable behavior. The laws of physics explain physical behavior. We must consider biological behavior in order to understand the nature of the biological principle. To understand the difference of biological principle from the physical one, we must compare the behavior of inanimate objects and living organisms.

*Philip Ball*
It is clear that organisms possess 'motivations', which means that their trajectories, while not violating Newtonian mechanics, can't easily be deduced from that.

*Attila Grandpierre*
It is not only the case that biological behavior "can't easily be deduced from Newtonian mechanics". The plain fact is that the most action principle cannot be derived from the least action principle, independently of trying easily or more systematically. The first principle of biology simply cannot be derived from the physical principle. The fundamental complexity measures I considered in my chapter underpin this fact quantitatively. Since the algorithmic complexity cannot be produced by physical processes, and because the laws of physics has an algorithmic complexity around 1 000 bits (see in my chapter), therefore the algorithmic complexity of biological organisms, being much higher than the 1 000 bits of physical laws, cannot be derived from physics.

*Philip Ball*
But I'm not clear why this should be a mystery that requires any principles beyond the ones we have already.

*Attila Grandpierre*
On the contrary. Biological behavior is a mystery at present only when narrowing down our considerations to the framework of physics. This apparent mystery can be resolved on the basis of a principle which is published more than eighty years ago by Ervin

>Bauer, which is, unfortunately, still ignored. So we do not require any principles beyond the ones we have already.
>
>*Philip Ball*
>Why doesn't evolution alone suffice to provide the imperatives and mechanisms, which are then acted out in particular situations and contexts by particular organisms?
>
>*Attila Grandpierre*
>The evolution of species is merely a historical process of special, namely, earthly life forms and it should not be taken as the most fundamental biological process. The most fundamental biological processes are the ones corresponding to the most action principle, which are the virtual interactions generating biological couplings between exergonic, energy liberating and endergonic, energy requiring processes. Evolution is like the growth of the tree. Similarly, the growth of the tree is not the fundamental life process of the tree. Cellular biochemical activity, the biological couplings governed by the most action principle, and the virtual interactions manifesting the most action principle are more fundamental biological processes. Growth of the tree is the result of cellular activity and not the other way around. One cannot explain cellular biochemical reactions in terms of growth of the tree. The "biological imperatives" arise ultimately not from evolution and selection pressure, but from biological organization governed by the Bauer principle.

(continued)

*Philip Ball*
Grandpierre argues that abiogenesis cannot seem to create, in a sufficiently short time, the complexity we see in life: if I understand correctly, he implies that only life (or 'intelligence') can beget life. To my mind, there are two shortcomings with this. First, it assumes that accumulation of complexity is linear, whereas it now seems that many complex systems possess thresholds above which entirely new modes of behaviour – new capabilities – appear.

*Attila Grandpierre*
Actually, I argued in my chapter that the accumulation of complexity can be faster in living organisms than in abiotic systems. Moreover, one of the main results of my paper is that I determined quantitatively the rate of complexity upjumps from abiotic systems to the first simple living organisms, and the complexity upjumps from the smallest bacteria to humans. The results tell that the complexity upjump assumed in the hypothetical abiogenesis during a period less than hundred million years is larger than the complexity upjump during the four billion years of biological evolution. I argued that this result makes the assumption of abiogenesis improbable.

*Philip Ball*
As I understand it, this is consistent with my claim that, once you have a bacterium, all the rest (up to humans) follows. I agree that the origin of life is a huge step in increase of complexity. But I'm not sure that we should necessarily have any expectations about whether life, once begun, would maintain a comparable rate of increase in complexity. Bacteria and other single-celled organisms are extremely successful; humans are anomalies. Work like that of Stuart Kauffman at least claims to show that, once you have autocatalytic cycles, you have the basic ingredients of 'life' in place.

*Attila Grandpierre*
Autocatalytic cycles when coupled together into an integrated unit form an automaton. Organisms are much more than automata since they are able to reorganize themselves by biological organization. It follows that biological organization is not the result of autocatalytic cycles but is the cause of organizing such cycles, together with ingenious reactions, into a living organism. Indeed, bacterial autonomy, motivations and significant achievements (see the references indicated above) show the profound significance of non-mechanically repeated biological reactions.

*Philip Ball*
In any event, I'm not sure we understand the processes that lead to a living organism well enough yet to be able to make convincing claims about how 'unlikely' abiogenesis is.

*Attila Grandpierre*
The argument presented in my chapter is a quantitative one which is more close to the norms of established science than opinions which did not reach the phase of quantitative arguments.

*Philip Ball*
Hence my comment below: it seems you are introducing a new idea/process largely because we don't yet have enough understanding to bridge the gap on the basis of established ideas, rather than because there is any clear empirical demand for it.

*Attila Grandpierre*
There are arguments indicating that complexity science is the next frontier of natural sciences as well as arguments telling that the $21^{st}$ century is the century of biology. Unfortunately, it seems that both complexity science and theoretical biology lacks its exact base which could be comparable to the exact base of physics. In my chapter I tried to introduce fundamental measures of complexity. I found that besides the complexity we observe with our eyes (morphologic or phenomenal complexity) there exist a fundamentally different level of complexity which is already known as algorithmic complexity. I tried to show that genetic complexity, which is a static thing, represents a still deeper level of complexity. I argued that biological organization has a complexity corresponding to the genetic level of complexity, but, at variance with genetic complexity characterized by the number of non-coding base pairs, biological organization represents a form of activity. With the help of these fundamental complexity measures I tried to obtain new light on the difference between machines and organisms. I hope I succeeded to make some steps towards a quantitative theory of biological organization and show the possibility of putting complexity sciences and theoretical biology to a more firm basis. Actually, there is a whole list of experimental data and theoretical underpinnings indicating the increasingly vital need for a realistic complexity science and theoretical biology (given in Grandpierre, 2007, *NeuroQuantology* 5, 346-362.). Therefore, I did not introduce a mystic new factor to fill the gaps of our present knowledge. Instead, my approach explains biological behavior on the most fundamental, exact and elegant way possible, since the most action principle of biology and the least action principle of physics form a natural union: the general action principle.

*Philip Ball*
To my mind, "intelligence in Nature" here becomes another God of the gaps.

*Attila Grandpierre*
To my mind, the first principle of physics is not another God of the gaps. It is quantitative, predictive, and consistent with our best and broadest theories and the widest range of established empirical facts. Similarly, the biological first principle is also quantitative, predictive, it is the best and broadest biological theory, consistent with the theoretical biology of Ervin Bauer, and with a large body of yet unexplained facts that it can explain. This is science and not religion. In the twenty-first century more and more biological data are accumulated. In the absence of a general theoretical biology, there is an increasing frustration between millions of biologists (Brent and Bruck, 2006, *Nature* 440, 416). Recently, biological physics became a new frontier of natural sciences (Phillips and Quake 2006, *Physics Today* 2006; 59(5): 38-43.; Sung, 2006, *Crossroads: Journal of Asia Pacific Center for Theoretical Physics* 2006; 4: 1-3.). Biological physics is the interdisciplinary effort to cross the barriers between physics and biology from the biology side (Sung, 2006). The US National Science Foundation allocates billions of dollars year by year for initiating the birth of theoretical biological physics (Ladik, 2004, *Journal of Molecular Structure* 2004; 673: 59-64.). "Promoting research that encourages a holistic perspective to "understand complex systems" is a long-term investment priority in the strategic plan of the National Science Foundation of the United States" (Hübler, 2007, *Complexity* 12, No. 5, 9). The fundamental complexity measures of nature introduced in our chapter serves both needs simultaneously.

I think it is time to realize the need for a realistic theoretical biology and complexity science.

*Philip Ball*
**"**I think it is time to realize the need for a realistic theoretical biology and complexity science." I completely agree!